# HUMAN-ARTIFICIAL INTELLIGENCE TEAMING FOR SCIENTIFIC INFORMATION EXTRACTION FROM DATA-DRIVEN ADDITIVE MANUFACTURING RESEARCH USING LARGE LANGUAGE MODELS

**Mutahar Safdar**[†]
McGill University
Montreal, QC,
Canada

**Jiarui Xie**[†]
McGill University
Montreal, QC,
Canada

**Andrei Mircea**[†]
University of Montreal
& Mila
Montreal, QC,
Canada

**Yaoyao Fiona Zhao***
McGill University
Montreal, QC,
Canada

[†] Equal Contribution, *Corresponding Author

## ABSTRACT

*Data-driven research in Additive Manufacturing (AM) has gained significant success in recent years. This has led to a plethora of scientific literature to emerge. The knowledge in these works consists of AM and Artificial Intelligence (AI) contexts that haven't been mined and formalized in an integrated way. It requires substantial effort and time to extract scientific information from these works. AM domain experts have contributed over two dozen review papers to summarize these works. However, information specific to AM and AI contexts still requires manual effort to extract. The recent success of foundation models such as BERT (Bidirectional Encoder Representations for Transformers) or GPT (Generative Pre-trained Transformers) on textual data has opened the possibility of expediting scientific information extraction. We propose a framework that enables collaboration between AM and AI experts to continuously extract scientific information from data-driven AM literature. A demonstration tool is implemented based on the proposed framework and a case study is conducted to extract information relevant to the datasets, modeling, sensing, and AM system categories. We show the ability of LLMs (Large Language Models) to expedite the extraction of relevant information from data-driven AM literature. In the future, the framework can be used to extract information from the broader design and manufacturing literature in the engineering discipline.*

## 1. INTRODUCTION

Additive manufacturing (AM), commonly known as 3D printing, fabricates parts layer-by-layer [1]. Offering unique benefits, the process can rival conventional manufacturing techniques. This has inspired significant research efforts into the technology aimed at enhancing its maturity for industrial adoption. A major portion of the research from recent years has relied on machine learning (ML) or deep learning (DL)-based approaches following the success of advanced data analytics techniques [2, 3]. The scientific works at the intersection of two growing disciplines are extremely information-rich. It is critical to extract relevant information from the incoming literature flux in order to reproduce and adapt these solutions for real-world applications.

The plethora of emerging literature has led to several state-of-the-art reviews to summarize the development and highlight the future of technology [1-3]. These reviews are divided across process technologies, applications, and solution types with varying scopes. This leads to subjective and time-irrelevant information being captured. The efforts to summarize data-driven AM research are limited in scope due to several reasons and fail to provide an all-encompassing reusable approach to information retrieval. Solutions capable of extracting the most relevant information from data-driven AM research are needed that can be re-used across a range of topics (e.g., technologies, applications, and data analytics solutions) in the field.

The challenge of extracting relevant information from science and engineering publications is not new and dates back to the 1960s [4]. Many scientific disciplines are faced with the high flux of newly published literature limiting access to relevant information [5]. As a result, Scientific Information Extraction, or SciIE is an established field though its maturity varies across disciplines. In general, Information Extraction or IE refers to a set of techniques in Natural Language Processing (NLP) that enable automated retrieval of structured information from text [6]. The solution to extract information can take many forms once the raw data is processed and cleaned. In their review on SciIE, Hong et al. identified vocabulary generation, text classification, named entity recognition, and relationship extraction as some of the steps in the information retrieval pipeline [7].

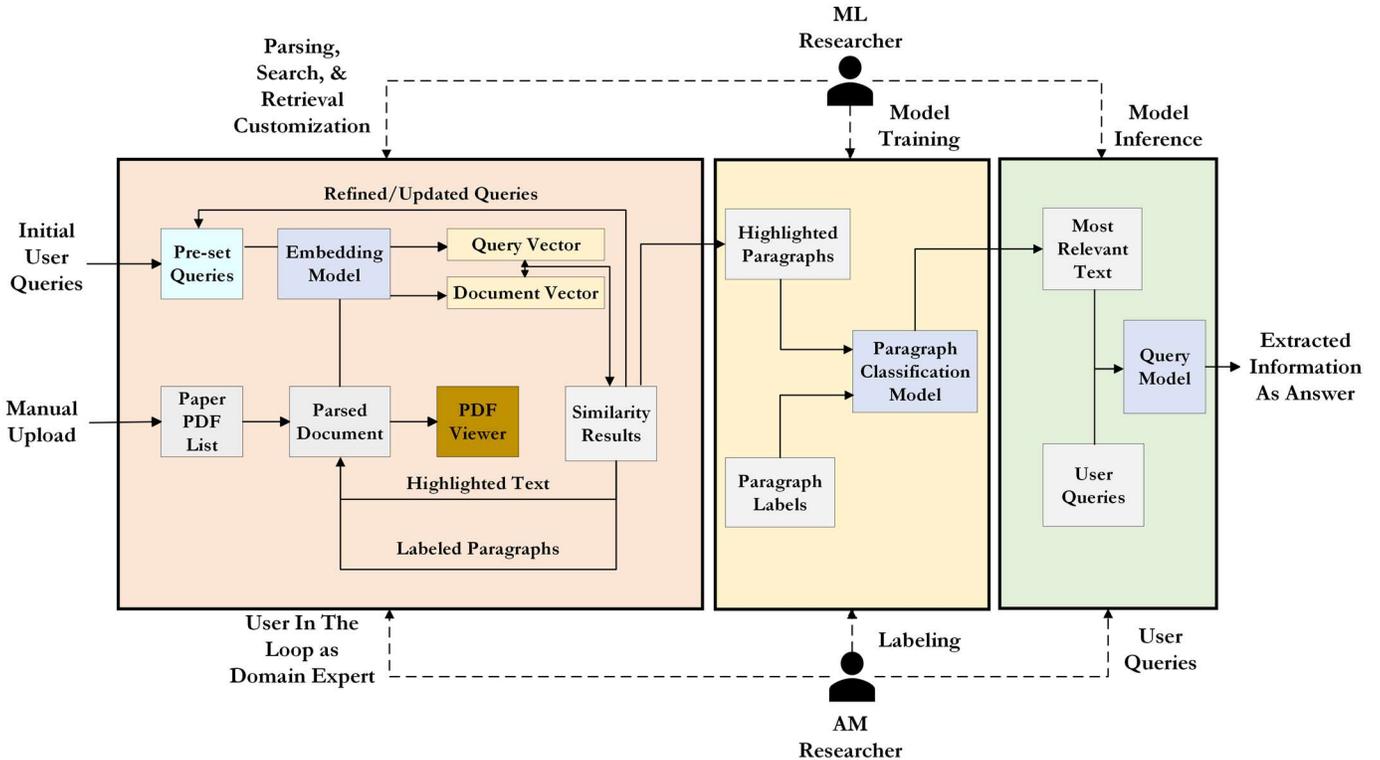

**FIGURE 1:** INFORMATION EXTRACTION FRAMEWORK WITH BASE IE SYSTEM (Left), PARAGRAPH CLASSIFICATION TIER (Middle), AND QUERY TIER (Right). THE SUBCOMPONENTS ARE EXPLAINED IN THE FRAMEWORK AND CASE STUDY SECTIONS

SciIE techniques can be broadly classified into traditional methods or more recent approaches based on ML or DL. In the past, structured information extraction in SciIE has relied on manual curation and data mining methods. This required domain experts to manually annotate and index research papers, which took a lot of time and resources [7]. However, recent advancements in NLP and ML have revolutionized the field. Modern approaches use DL models like transformers to automatically extract useful information from unstructured text [8]. These models can identify key entities, relationships, and contextual information from scientific documents, enabling rapid and scalable IE. Additionally, modern methods take advantage of domain expertise to improve the precision and relevance of information retrieved. This combination of traditional curation and cutting-edge technology has the potential to significantly accelerate scientific discovery and information retrieval in research domains such as data-driven AM.

Large language models or LLMs are based on transformer architectures (models designed for sequence-to-sequence tasks) and employ self-attention which enables the model to weigh the importance of different segments within the text [9]. Their introduction has been a breakthrough in AI and their capabilities make them well-suited for key NLP tasks including IE. While LLMs can support IE in various ways, their direct application in scientific disciplines may lead to lower performance. There are several reasons that can result in model hallucination such as domain specificity, limited context, ambiguity, outdated knowledge, low signal-to-noise ratio, and data privacy. There is a need to overcome this gap when leveraging powerful LLMs for SciIE.

To utilize LLMs for effective IE in data-driven AM, this work proposes a human (domain expert) centered approach integrating text retrieval, classification, and generation models. We particularly focus on making the extraction process transparent by having human review and feedback incorporated at each step. The remainder of the paper is as follows. Section 2 introduces the framework. Section 3 demonstrates the human-centered features. Section 4 details the case study and its steps. Section 5 presents the results and discussions. We conclude this article with concluding remarks and future works in Section 6.

## 2. INFORMATION EXTRACTION FRAMEWORK

The proposed IE framework is divided into three major components namely the base IE system, paragraph classification tier, and the query tier. The base IE system acts as the engine of the IE framework and allows AM and AI researchers to interact through a GUI to iteratively improve the answers coming from the query tier. Figure 1 presents an overview of the framework and outlines the key steps of each component. These components are explained in detail below.

### 2.1 Base IE System

The base IE system acts as the engine of our IE framework allowing users to upload scientific articles as PDF files, parsing the PDF files into paragraphs and equations for downstream retrieval, allowing AM experts to search relevant paragraphs through regular or semantic search, and customizing retrievals to effectively shortlist relevant paragraphs.

Based on the methodology of Lo et al., the base IE system processes user-provided PDFs of scientific AM articles with GROBID [10]. Metadata such as publication title, abstract, authors, date and DOI are first extracted. Then GROBID parses a paper's paragraphs and equations organized by section; as well as figures, tables and references (with corresponding links in the text). Lastly, the different parsing outputs are saved to the user's library for use in the IE system, along with the original PDFs for user reference. Despite its widespread use, GROBID still struggles with certain PDFs, displaying parsing errors like missing or duplicate paragraphs, invalid organization of sections, incorrect or missing metadata, and illegible equations. In alignment with our human-centered considerations, we ensure transparency and iterative improvement by letting users view and correct parsed results with the original PDFs for reference.

IE systems typically involve an information retrieval (IR) step, where relevant documents are first retrieved [11]. Because our use case involves extracting information from a given paper, we retrieve relevant paragraphs instead. Specifically, we generate embeddings for paragraphs and for queries associated with four categories of interest defined by the AM experts (e.g., data, modeling, sensing, and AM systems) We use cosine similarity to retrieve the paragraphs that are most similar to a given query. Because of computational constraints and maintainability concerns associated with hosting the IE system, we decided to use the OpenAI embeddings API instead of our own models.

While we could perform IE on the full text of a paper, this has significant drawbacks that motivate paragraph retrieval. First, for a given information item, typically only a small portion of the full text contains the desired information. Conversely, feeding full-texts into our LLM-based IE model is infeasible or too costly to be viable. Second, limiting our LLM-based IE model to retrieved paragraphs —which are typically quickly readable and self-coherent— enables rapid and easy user-verification of our IE system's output by cross-referencing with retrieved passages. As errors are inevitable in any IE system, and pernicious with LLMs, this human-centered consideration helps build trust through transparency (Explained in Section 3.2).

Similar to Dunn et al., we use LLMs to perform IE on scientific texts with a sequence-to-sequence formulation [12]. However, in the process of participatory design (Explained in Section 3.4), we found that rigidly structured outputs did not lend themselves well to extracting specific information items across multiple categories. Instead, we prompt the LLM to synthesize the retrieved passages, extracting the information that is relevant to the query of a given information item, or clearly indicating cases where no such information is present. As mentioned earlier, we use the OpenAI chat API instead of our own models due to constraints on computational resources and maintainability.

As seen in Figure 2, the query used to prompt the LLM-based IE model may or may not be appropriate for retrieving relevant paragraphs. Throughout development, this became rapidly apparent, and users began experimenting with different retrieval queries and building intuitions for common failure mechanisms and working mitigations. However, sharing queries between the IR and IE systems limits the flexibility afforded to users in designing a query, as it is constrained to being a valid instruction interpretable by the LLM-based IE system.

To address this limitation and enable iterative improvement (Explained in Section 3.2), we create an interface that allows users to create and update custom retrievals with ensembles of positive and negative queries Q+ and Q−, as well as positive and negative paragraphs P+ and P−. Positive queries are meant to retrieve similar paragraphs, while negative queries are meant to prevent retrieving similar paragraphs. Conversely, when retrieving paragraphs, users can annotate these as positive or negative to obtain a similar effect. The interfaces for these functionalities are shown in Figure 2 (B-E). Concretely, we compute a retrieval embedding R as a sum of the averages of these embeddings, weighted by a, b, c, d:

$$R = \sum_p^{P+} \frac{ap}{|P+|} + \sum_q^{Q+} \frac{bq}{|Q+|} - \sum_p^{P-} \frac{cp}{|P-|} - \sum_q^{Q-} \frac{dq}{|Q-|} \quad (1)$$

While simple, this human-in-the-loop approach enables users to improve retrieval in a way that is intuitive and enables fine-grained control or experimentation; all the while not interfering with or being constrained by the IE system query.

**2.2 Paragraph Classification Tier**

We expect to find the relevant information in specific paragraphs and hence there is value in training paragraph classifiers to further expedite the IE process. In the long term, classifiers specific to a certain domain can quickly filter the relevant paragraph from the whole article. These relevant paragraphs can then be used in the query tier to provide specific answers to the readers. It is also important to mention that currently the paragraph classification is done at an abstract level (e.g., data as compared to specific data characteristics or modeling as compared to specific modeling details) to use shallow and light-weight classifiers that simplify the training process.

One of the key goals at this stage is to prepare global classifier(s) for each field to quickly filter the relevant paragraph from irrelevant. The accuracy of the classifiers is expected to grow gradually as the domain experts read through the papers and label the paragraphs in the base IE system. As a result, this will incrementally decrease the effort of manually going through each article in the base IE system. Nonetheless, the long-term effectiveness of these classification models will depend on regular fine-tuning with newly labeled paragraphs. This could be inspired by the poor performance of the trained classifiers on certain components of scientific information.

$$l_{ij} = \sigma(w_j^T p_i + b_j) \quad (2)$$

In the proposed framework, multilabel l(i,j) paragraph (Pi) classification models (e.g., Equation 2) are being built offline on top of the data labeled in the base IE system. This is to provide flexibility in the choice of a model (shallow vs deep, binary, multi-class or multi-label) that works best for each scientific domain. The functionality to deploy trained models online by integrating them into the base IE system can be implemented in the future.

**2.3 Query Tier**

The question or query tier represents the last component of the proposed IE framework. It simply allows the users to ask a specific question and expect a well-formulated answer. For this purpose, the filtered paragraphs are fed into a GPT model along with the user query to output the answer. Currently, a query function is available in the base IE system to directly filter the information. In the future, we expect query functions to also interact with the outputs of classification models once these are integrated into the base IE system. Equation 3 represents the structure of the prompt where both query (Q) and paragraphs (Pi) are fed to the model while being kept apart through a separator (SEP).

$$GPT_i = Q + |SEP| + P_1 ... + |SEP| + P_n \quad (3)$$

**3. HUMAN-AI TEAMING FEATURES**

The Human-AI teaming is central to the proposed IE framework to keep both AM and AI researchers in the loop as the information retrieval pipelines specific to a certain domain are optimized. These are oftentimes referred to as human-centered considerations in the broader NLP literature. In the context of data discovery, a problem similar to IE, Gregory and Koesten refine the notion of "human-centered" as thinking from the perspective of the person(s) engaging in the activity; with a focus on the interaction process and the "user" experience, taking into account different contexts and needs [13]. In a different vein, Egan et al. present "user-centered" NLP systems as a human-computer collaboration where computers do what they do well (process large amounts of information, filter, sort and prioritize) and humans do what they do well (assess, select, and refine with domain expertise) [14]. And lastly, on a more abstract level, Kotnis et al. define "human-centric" NLP research as a process where human stakeholders actively participate in the research [15]. In this section, we discuss several human-centered considerations which relate to these formulations and have influenced the development of our IE system.

**3.1 User Interface**

A prototype tool is implemented that reflects the base IE system explained in the previous section. Figure 2 shows an overview of the tool. The tool provides several functionalities to the users including the option to create libraries to group PDF files from similar domains. Each library provides a list of papers and highlights the author and publication data. The uploaded PDF files can be viewed as-is. In addition, a simple text search or semantic search can be performed on the parsed PDF files.

The Query tab represents the option to write specific queries. The Retrieval tab highlights the functionality to create retrievals and iteratively update them by labeling the paragraphs as positive or negative. The user interface enables several human-centric features that are explained in the following subsections.

**3.2 Transparency and Trust**

IE systems ideally improve the efficiency of their users, however, Schleith et al. suggest that a lack of transparency can lead to a lack of trust [16]. This lack of trust can in turn undo efficiency gains as users spend more time carefully reviewing system outputs they do not trust. Similarly, but in the context of LLM-generated summaries, Cheng et al. present appropriate trust as enabling users to decide whether or not to rely on a given system output; which in turn requires transparency [17]. These challenges are all-the-more important for LLMs, which are known to generate convincing but erroneous confabulations.

We build our IE system around this human-centered consideration of trust by transparency in a variety of ways. First, as noted in the base IE system, errors are introduced as early as the data parsing stage. By overlaying interactions with the IE system on top of this raw data, we enable users to more easily catch and correct errors related to parsing, building appropriate trust. Additionally, transparency on the data level is essential for mitigating potential issues of data cascades [18].

As mentioned in Section 2.1, we also introduce an intermediate retrieval step before performing LLM-based IE. While this can improve factual consistency, it does not completely prevent confabulation [19]. However, adapting the interface to enable streamlined cross-referencing of system outputs with retrieved passages (specifically paragraphs, to facilitate quick verification) builds appropriate trust where eliminating errors is otherwise infeasible.

**3.3 Iterative Improvement**

While transparency builds appropriate trust by supporting users in deciding whether to rely on system outputs or not, iterative improvements in IE systems can minimize the rate at which users should decide not to rely on a given output, improving their experience. More specifically, human feedback with human-in-the-loop approaches can be leveraged to improve the reliability of IE system outputs [20]. We adapt this human-centered consideration by enabling users to create custom retrievals where they can provide feedback on retrieved passages and iteratively improve the reliability of the custom retrieval. More generally, Rahman and Kandogan find that human-in-the-loop IE workflows are typically iterative in nature, characterized by information foraging and sensemaking loops as users iteratively improve their understanding of the task and the data [21]. We try to support this consideration and give users the flexibility required for this kind of iteration. Notably, we give users fine-grained control over the underlying retrieval and extraction systems so they can experiment with different approaches.

**FIGURE 2:** OVERVIEW OF THE PROTOTYPE TOOL IMPLEMENTED. (A) REPRESENTS THE MAIN VIEW OF THE TOOL IN A WEB BROWSER ONCE THE IE VALIDATION CHECKLIST OF 100 PAPERS IS SELECTED. THE PAPERS ARE LISTED ON THE LEFT SIDE WHEREAS THE PDF OF THE SELECTED PAPER IS SHOWN ON THE RIGHT. (B) REPRESENTS THE STRING-BASED TEXT SEARCH FUNCTION. (C) REPRESENTS THE SEMANTIC SEARCH WHICH VECTORIZES USER QUERY FOR COSINE SIMILARITY. THE RESULTING TOP FIVE PARAGRAPHS ARE HIGHLIGHTED. THE ARROWS NEXT TO THE SEARCH BUTTON ENABLE USER VERIFICATION OF THE OUTPUT WITH EASY NAVIGATION BETWEEN RETRIEVED PARAGRAPHS USED BY THE IE SYSTEM. THE HUE OF THE HIGHLIGHT INDICATES THE STRENGTH OF THE COSINE SIMILARITY BETWEEN THE QUERY AND THE PASSAGE. (D) REPRENTS THE QUERY FUNCTION AND THE RESULTING ANSWER GENERATED FROM TOP FIVE RELEVANT PARAGRAPHS. (E) REPRESENTS THE FUNCTIONALITY TO CREATE, SELECT AND UPDATE A RETRIEVAL

### 3.4 Participatory Design

An important consideration throughout this work has been participatory design: ensuring a relationship of meaningful co-creation and mutual learning between users and researchers [22, 23]. This collaborative approach between developers (machine learning researchers) and users (mechanical engineering researchers) enabled iterative refinements throughout the development of the IE system prototype; from initial brainstorming and problem formulation to interface adaptations that address user-identified limitations of the underlying machine learning models. Crucially, our approach is fundamentally human-in-the-loop rather than human-on-the-loop. In other words, users and researchers actively participate in a task rather than passively supervising or validating its automated completion. We found this dynamic played a significant role in fostering participatory design throughout development.

### 4. CASE STUDY

Inspired by the increasing scientific publications as shown in Figure 3, a case study was conducted using literature at the intersection of AM and ML. It is particularly challenging to find key components of information quickly and effectively from literature at the intersection of two growing fields. The tool enables an interactive way to query the information required and hence provides an opportunity to go through the literature quickly as compared to relying on existing reviews. The reviews become outdated with time and are limited in the way information can be represented. In addition to providing a faster and effective way to retrieve key information components, the tool can be used for other domains and applications so as to provide a reproducible pipeline for SciIE.

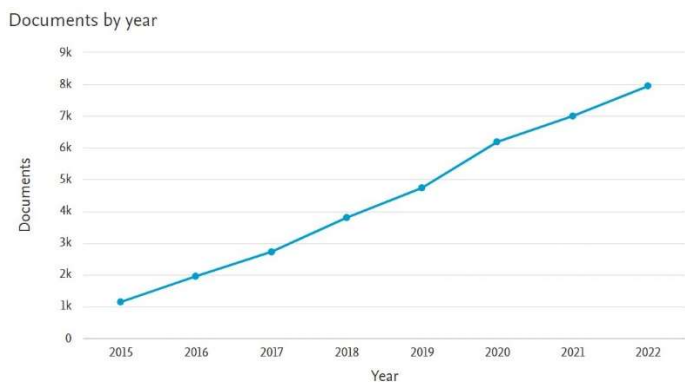

**FIGURE 3:** THE PLOT HIGHLIGHTS THE INCREASING LITERATURE IN ADDITIVE MANUFACTURING INSPIRING THE CREATION OF AN AI TOOL TO QUICKLY AND EFFECTIVELY FILTER KEY SCIENTIFIC INFORMATION.

### 4.1 Defining Relevant AM+AI Information

We categorized the information contained within the data-driven AM literature into four categories which jointly represent most of the key information required to understand and evaluate the presented research. These categories are listed below:

- *Data Relevant*: Information representing data for ML applications such as data characteristics, experimental settings, data preparation, data processing, and data availability [24, 25].
- *Model Relevant*: Information related to ML-based modeling such as the algorithm, training process, the compute hardware & software, and model availability.
- *Sensing Relevant*: Information relevant to sensing technique and equipment such as the physical phenomenon, sensor type, sensor specifications, sensor settings, and sensor deployment.
- *System Relevant*: Information representing manufacturing technology, hardware, and materials used

### 4.2 Collecting Research Articles

In order to conduct the case study, we retrieved 100 research articles representing ML-based research on in-process monitoring and quality prediction challenges in AM. The latest publication year among the articles is 2023 whereas no limit was set on the starting year. These papers represent a diverse and comprehensive body of research in ML-driven AM research. All articles were collected from Scopus. The decision to use 100 research articles to validate the IE pipeline was made to get a representative dataset spanning various subdomains in AM and ML. The PDF files of all articles were downloaded and grouped into an "IE Validation" library inside the prototype tool. As soon as the PDF of an article is uploaded, it is parsed at the backend to support subsequent search, labeling, and retrieval.

### 4.3 Searching, Labeling and Retrieval Customization

Once the PDF files were added in the prototype tool, these were parsed to act as the input of the embedding model. The current version of the prototype tool used the OpenAI text-embedding-ada-002 model to provide a paragraph-level vector representation of the parsed PDF files. This set the stage to find relevant information through semantic search where input query was also featurized using the same model and the two were compared using cosine similarity. However, in the case study, we relied primarily on the retrieval functionality (Figure 4) and iteratively updated it by selecting positive and negative paragraphs to compute the retrieval embedding of Equation 1. As will be shown in the results, the ranking results and score gradually improved as we went through the papers labeling the paragraphs. This reflected the effectiveness of customizing the retrievals specific to each relevant information category.

Where specific information was not found in the top highlighted paragraphs from customized retrievals, we used search functionality to find it. If the relevant paragraphs were found through the search functionality, they were labeled as positive to include them in the retrieval embedding. Similarly, where irrelevant paragraphs were found in the top results of a specific retrieval, these were marked as negative to be excluded from future results. Figure 5 shows the relevant and irrelevant paragraphs highlighted as positive and negative to account their respective embeddings in the overall retrieval embedding. The labeling process led to a multi-label text dataset for ML-driven AM literature to be used in the next step. To the best of the author's knowledge, this is the first NLP dataset in AM [24, 26].

labeled into four relevant categories namely data, model, sensing, and system. We introduced a fifth category for paragraphs that didn't belong to any of the above-mentioned categories as "irrelevant." This was done to evaluate the effect of including these paragraphs in the learning process. However, their inclusion introduced data imbalance, and these were subsequently removed as a redundant category label.

The dataset was processed using the OpenAI embedding model to generate feature vectors for each paragraph. The augmented dataset was used to train a Random Forest classifier from the Scikit-learn library. Since the classifier doesn't natively support multi-target classification, we used the built-in MultiOutputClassifier strategy. We trained a simple multi-label model as a global classifier for ML-based AM literature. The classifier can categorize the paragraph into four categories of relevant information. However, irrelevant paragraphs will require to be filtered and out-of-balanced classes should be down-sampled for future training of the model. The results are presented in the next Section.

### 4.5 Query and Response

During the case study, we used a query function that prompts an LLM to extract the relevant information by providing both a user query and the relevant paragraph(s) to generate the answer. This functionality can be used both during the labeling process to find missing information as well as after the classifiers have been trained to filter the relevant paragraphs. Table 1 shows the function used to prompt.

**TABLE 1:** FUNCTION TO PROMPT LLM AND SUPPORT SPECIFIC IE RETRIEVAL

| |
|---|
| **Function**: create_prompt |
| **Description**: return the required information from the relevant passages |
| ```def create_prompt(query: str, retrieved: list[str]):
    retrieved = [f"- Passage {i}: {x}" for i, x in enumerate(retrieved)]
    retrieved = "\n".join(retrieved)
    prompt = f""" You are an assistant for a researcher working at the intersection of additive manufacturing and machine learning. Your goal is to help the researcher find and distill significant information in a scientific paper. To this end, answer the following triple-backtick delimited query from 404 the researcher:
``` {query} ```
To answer the question, use the following passages from the paper. If there is no information in the passages that answers the question, write "I cannot answer that."
{retrieved}
"""
    return prompt``` |

## 5. RESULTS AND DISCUSSIONS

The results from the case study are divided into two categories. Table 2 shows the improvement in the ranking of

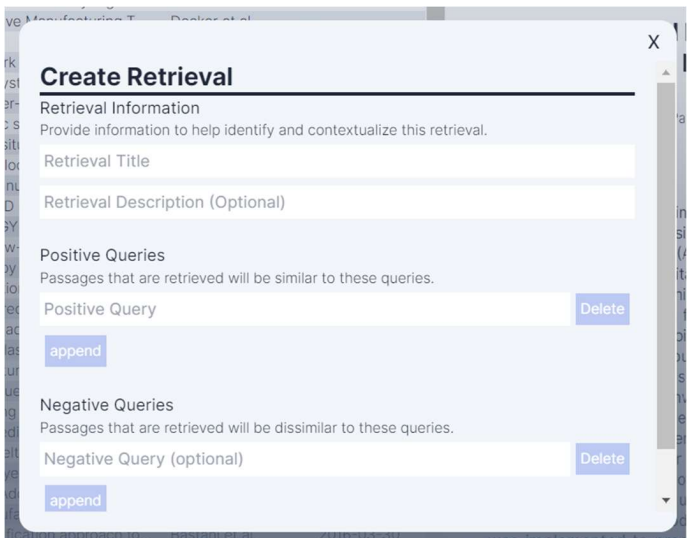

**FIGURE 4:** RETRIEVAL CREATION WINDOW. A RETRIEVAL CAN BE NAMED AND POPULATED WITH POSITIVE (RELEVANT E.G., DATA, MODEL) AND NEGATIVE QUERIES (IRRELEVANT E.G., RELATED WORKS SECTION). WE CREATED FOUR RETRIEVALS AS SHOWN IN THE APPENDIX.

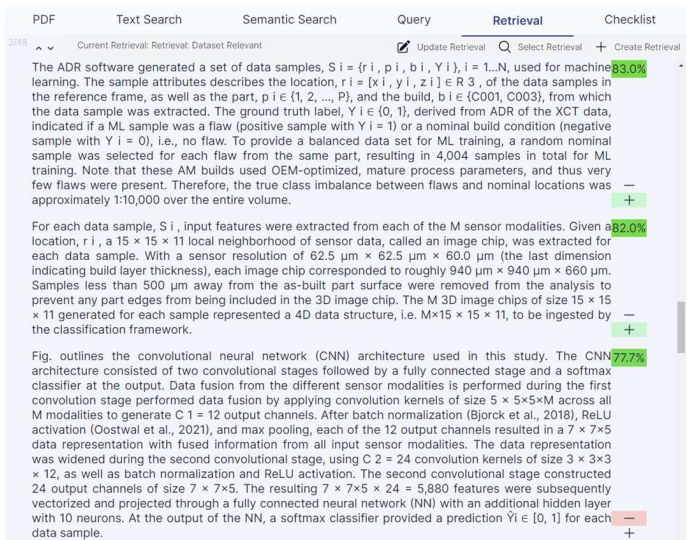

**FIGURE 5:** A POPULATED RETRIEVAL (DATA RELEVANT IN THIS CASE) CAN BE CUSTOMIZED BY SELECTING THE RELEVANT AND IRRELEVANT PARAGRAPHS. BY CLICKING ON A PLUS SIGN BUTTON, PARAGRAPHS CAN BE ADDED TO P+ TO INCREASE THE CHANCE OF RETRIEVING SIMILAR PARAGRAPHS FOR A GIVEN CUSTOMIZED RETRIEVAL. WE CAN ALSO ADD PARAGRAPHS TO P− TO REDUCE THE CHANCE OF RETRIEVING SIMILAR PARAGRAPHS FOR THE SAME CUSTOMIZED RETRIEVAL.

### 4.4 Classifying Paragraphs

The resulting multi-label paragraph dataset was downloaded from the prototype tool and used to develop paragraph classifiers as global domain models to rapidly filter relevant paragraphs for downstream IE. The raw dataset represented 5039 paragraphs

**TABLE 2:** IMPROVEMENT IN SIMILARITY RANKING FROM FIRST TO LAST PAPER IN THE CASE STUDY. THE COMPARISON IS DONE BY SELECTING A RANDOM PAPER AND APPLYING DEFAULT RETRIEVALS. THE RESULTS FROM DEFAULT RETRIEVALS AND FINAL CUSTOMIZED RETRIEVALS ARE COMPARED. AS EVIDENT FROM THE RESULTS, THE INITIAL RANKING LEADS TO IRRELEVANT PARAGRAPHS WITH VARYING SCORES WHERAS THE CUSTOMIZED RETRIEVALS LEAD TO TOP RELEVANT PARAGRAPGH WITH REQUIRED INFORMATION

| IE Category | Initial Ranking | Final Ranking |
|---|---|---|
| Data Relevant | In the experiment, there were two main difficulties when applying the proposed method. For the experimental setup, the camera location and settlement were difficult for the series of track image sampling although the focal length and the sampling angle of the camera were fixed, which affected the quality and acquisition of the images. The optimal location of camera needs to be further investigated. For the detection approach, the proposed DBN structure is still in the stage of in situ SLM melted state recognition. This means that it works well when images are captured online and data is processed offline. The system setup and the algorithm will be improved to raise the recognition speed and applied for the real-time monitoring in the future work. — 83.3% + | With the sampled images, an area of 400-pixel (columns) by 500- pixel (rows) was taken out for covering the area of interest. 100-pixel (columns) by 125-pixel (rows) image was obtained by down sampling for decreasing the computing memory. With statistical analysis, impulse noise existed among the images. A median filter was used to remove the non-Gaussian noise in order to obtain accurate knowledge from the monitoring system. The classification was to identify five kinds of melted states. 5000 frames of each A to E pattern were used for training, and 800 frames of each experiment 1 to 5 were randomly selected for testing. — 81.0% + |
| Model Relevant | In the experiment, there were two main difficulties when applying the proposed method. For the experimental setup, the camera location and settlement were difficult for the series of track image sampling although the focal length and the sampling angle of the camera were fixed, which affected the quality and acquisition of the images. The optimal location of camera needs to be further investigated. For the detection approach, the proposed DBN structure is still in the stage of in situ SLM melted state recognition. This means that it works well when images are captured online and data is processed offline. The system setup and the algorithm will be improved to raise the recognition speed and applied for the real-time monitoring in the future work. — 83.8% + | Classification procedure was produced by a supreme DBN structure with four-level hidden RBMs. Each level had 400 units to map the input data. Numbers of pre-training epochs were all 50 with an unsupervised learning rate of 0.001 and 100 batch size. The number of fine-tuning epochs was 2000 with a supervised learning rate of 0.001 and 100 batch size. With the CNN model, the optimal model was built with four convolutional layers of × × × × and 5 5,3 3,3 3, 3 3 filters. Each filter had stride of 1 and zero padding after parameter selecting. Two max pooling layers took filters of × 2 2 with a stride of the same length with the convolutional layers. The mini batch size was 100 and the training has 20 epochs. With the extracted features, an optimal MLP model with ten hidden layers was established with 20 neurons at each hidden layer. A three-level DBN structure with 20 units for each RBM layer was suitable. The epochs of the MLP training and the DBN fine-tuning were both 1000 and learning rates were 0.001 for iteration. — 85.2% + |
| Sensing Relevant | In the experiment, there were two main difficulties when applying the proposed method. For the experimental setup, the camera location and settlement were difficult for the series of track image sampling although the focal length and the sampling angle of the camera were fixed, which affected the quality and acquisition of the images. The optimal location of camera needs to be further investigated. For the detection approach, the proposed DBN structure is still in the stage of in situ SLM melted state recognition. This means that it works well when images are captured online and data is processed offline. The system setup and the algorithm will be improved to raise the recognition speed and applied for the real-time monitoring in the future work. — 82.0% + | As shown in Fig. 1, the experimental monitoring system consists of a customized designed SLM system and a high-speed NIR camera. The SLM system was composed by a fiber laser with a beam diameter of 100 μm, a powder spreading device, and a controller panel. The controller panel could adjust the process parameters by connecting to the powder spreading device, laser power, and gas cylinder. The high-speed camera was FASTCAM Mini UX50/100 using sampling speed of 5000 fps at resolution 1024 × 1024 pixels with an angle of 30°over the platform. The images were captured with an external band-pass filter of 700 to 1000 nm. A series of SLM experiments was conducted on the 304 L stainless steel with the average diameter of 27 μm using an IPG YLR- 200-SM laser with 1064 nm wavelength. — 82.8% + |
| System Relevant | Insufficient energy density gives rise to a weak flow ability and a small contact area to the substrate, leading to balling phenomenon and high surface roughness [5]. Only the optimum building strategy and process parameter combination can overcome the defect caused by over or under melting. Numerous studies have been conducted to investigate the complex relationship between the defects and the melting process. These studies were carried to find out the relationship between the melting parameters and process phenomena incurred. For example, the relations between the process parameters and the characteristics of the melt pool [6], the plume [7], or the spatter [8] were investigated. The measurements by lens characteristics using the on-axis monitoring system will decrease the signal quality much, and the sensor setup is also an obstacle to such systems. Therefore, it is more suitable to collect the intense plume and spatter signatures by an off-axis monitoring system to detect the quality of parts. — 83.0% + | Additive manufacturing (AM), or 3D printing technology, has shown great advantages over conventional manufacturing processes due to the advancement in processing complex geometric, multi-material and multi-functional structures. The AM technologies have great potential in forming complex metal components for aerospace and healthcare applications. Among them, the selective laser melting (SLM) technology stands out as one of the most important metal AM processes, being able to produce free form geometrical parts in high quality and greatly reduce lead time [1]. However, concerns remain about the reliability and repeatability of the mechanical properties of finished parts. Issues such as irregular porosity, high surface roughness, cracks, and delamination continue to present challenges in the SLM process [2], and ideas of in situ process monitoring have been raised to solve these problems [3]. For this end, this paper proposed a convenient method to monitor and recognize the melted sates during the SLM process. — 82.7% + |

similarity across the four information categories. The passages are selected from one random paper out of the 100 and its most relevant paragraphs for four categories are shown against starting/default retrievals as well the customized retrievals after the AM researchers went through the 100 selected papers. The improvement in the similarity ensures that top matches to a selected retrieval contain relevant paragraphs with the required information. Appendix A shows the default retrievals associated with four information categories representing AM+ML literature that were used.

The second set of results represents the performance of the downstream global classifier on the multi-label paragraph dataset. The classifier was trained on a diverse set of ML-driven AM literature and its ability to classify represents initial success in building the global classifier. Table 3 represents the test performance of the Random Forest classifier in terms of precision, recall, and F1 score.

**TABLE 3:** CLASSIFICATION REPORT OF THE RANDOM FOREST MODEL ON TEST SPLIT OF PROCESSED AND RELEVANT PARAGRAPH DATASET. 0, 1,2,3 CORRESPOND TO DATA, SENSING, MODEL, AND SYSTEM CLASSES

| Class | Precision | Recall | F1-Score | Support |
|---|---|---|---|---|
| 0 | 0.85 | 0.79 | 0.82 | 121 |
| 1 | 0.84 | 0.89 | 0.87 | 110 |
| 2 | 0.88 | 0.90 | 0.89 | 101 |
| 3 | 0.86 | 0.90 | 0.88 | 49 |
| Micro avg | 0.86 | 0.86 | 0.86 | 381 |
| Macro avg | 0.86 | 0.87 | 0.86 | 381 |
| Weighted avg | 0.86 | 0.86 | 0.86 | 381 |
| Samples avg | 0.84 | 0.86 | 0.83 | 381 |

The similarity percentage fluctuates across paragraphs and is sensitive to their length. Moreover, the current approach lacks a threshold to define "good enough" similarity for relevant paragraphs. Similarly, the developed classifier is relatively simple since the post-process dataset is small as compared to those used in deep learning models of textual data. A higher-capacity global classifier requires more labeled data. For the next 100 papers, authors expect the effort to be significantly lower due to already customized retrievals. The options to generate synthetic data can be considered as well. In addition to increasing the data quality and quantity, irrelevant paragraphs need to be considered to further refine classifier boundaries.

## 6. CONCLUSIONS AND FUTURE WORKS

Inspired by the increasing frequency of research into data-driven solutions of AM challenges, we propose an information extraction framework powered by LLMs and built around human-centered considerations. The framework is divided into three components namely base IE system, classification tier, and the query tier whereas the query functionality is also integrated into the base system. The tool enables continuous update of the database representing a specific scientific domain while allowing domain experts to iteratively customize retrieval for LLM-based IE. The tool is deployed on the web and has restricted development access at the moment. We carried out a case study by building a library of 100 ML-based AM research articles and going through them to manually label and validate paragraph containing key information belonging to four categories namely data, model, sensing and system. We confirm the gradual effectiveness of customizing retrievals as we progress through the database. Moreover, the relevant paragraphs labeled as a result of retrieval customization were downloaded and used to train a shallow multi-label classifier. The results of classification on the test set indicate that it is possible to develop a global classifier for a given domain thereby significantly expediting the information filtering step.

The future works include:
- Validating the prototype tool and proposed framework in another design and manufacturing subdomain.
- Defining methods and metrics to benchmark IE efficiency and effectiveness as compared to existing tools and approaches.
- Introducing a notion of similarity threshold for relevant paragraphs in design and manufacturing scientific literature.
- Opening the tool to the broader design and manufacturing community to gather feedback.

## ACKNOWLEDGEMENTS

McGill Engineering Doctoral Award (MEDA) fellowship for Mutahar Safdar is acknowledged with gratitude. Mutahar Safdar also received financial support from National Research Council of Canada (Grant# NRC INT-015-1). McGill Graduate Excellence Award (Grant# 00157), Mitacs Accelerate Program (Grant# IT13369), and MEDA fellowship for Jiarui Xie are acknowledged with gratitude. The authors are grateful to Digital Research Alliance of Canada (RRG# 4294) for providing computational resources to support this research.

## DEMO OF PROTOTYPE TOOL

https://www.youtube.com/watch?v=gM7rFLmJEH0

# APPENDIX

### A-1: DEFAULT POSITIVE QUERIES OF DATA RELEVANT RETREIVAL

**Retrieval: Dataset Relevant**

Positive Queries:
- Are the data preprocessing techniques discussed, if any?
- The statistics of the raw and the final datasets such as data type, d...
- Are the experimental design settings for data acquisition introduced...
- Are the data preparation techniques discussed, if any?
- Are the relevant dataset statistics provided?

### A-2: DEFAULT POSITIVE QUERIES OF MODEL RELEVANT RETREIVAL (SHOWN PARTIALLY)

**Retrieval: Model Relevant**

Positive Queries:
- Are the details of model training described?
- If applicable, are hyperparameter search and selection procedures...
- The description of the type of ML algorithm. The description of the...
- Is the code open access or provided?
- Is the final trained model shared?

### A-3: DEFAULT POSITIVE QUERIES OF SENSING RELEVANT RETREIVAL

**Retrieval: Sensing Relevant**

Positive Queries:
- Are the sensor deployment details provided?
- Are the physical phenomena being captured by the sensing system...
- Are the sensor settings or calibration details provided?
- This question aims to ensure that the usage of the sensors is justifi...
- Are sensor type(s) and specification(s) provided?

### A-4: DEFAULT POSITIVE QUERIES OF SYSTEM RELEVANT RETREIVAL

**Retrieval: System Relevant**

Positive Queries:
- Is the base additive manufacturing system reported?
- Are the details on material system and material characteristic provi...
- Base additive manufacturing systems refers to the original off-the-s...
- In the case of a customized additive manufacturing system, are det...